\documentclass[aps,prb,twocolumn,showpacs,showkeys, floatfix]{revtex4-1}

\usepackage{graphicx}
\usepackage{dcolumn}
\usepackage{bm}
\usepackage{amsmath,amssymb,amsthm}
\usepackage{braket}
\usepackage{color}
\usepackage[english]{babel}

\begin{document}

\title{Spontaneous symmetry breaking in the laser transition}

\author{P. Gartner}
\affiliation{Centre International de Formation et de Recherche Avanc\'ees en 
Physique - National Institute of Materials Physics, Bucharest-M\u{a}gurele, 
Romania}

\email{gartner2@gmail.com}

\date{\today}

\begin{abstract}
In analogy with equilibrium phase transitions, we address the problem of the 
instability to symmetry-breaking perturbations of systems undergoing a laser 
transition. The symmetry in question is the $U(1)$ invariance with respect to 
a phase factor, and the perturbation is a coherent field $E$, coupled to the 
exciton.  
At the rate equation level we analyze first the case of a cavity containing a 
single, two-level emitter, and then a chain of such cavities interacting by 
photon hopping processes. In both cases spontaneous symmetry breaking takes 
place when the system is in the lasing phase. For the laser transition, the 
analogue of the thermodynamic limit is the scaling limit of vanishing cavity 
loss and light-matter coupling, $\kappa \to 0$, $g \to 0$, so that $g^2/\kappa$ 
remains finite. We show that in the lasing regime anomalous averages persist in 
the $E \to 0$ limit, provided that the scaling limit is performed first. Lasing 
diagnosis based on robust anomalous averages is compared numerically with the 
familiar coherence criterion $g^{(2)}(0)=1$, and the advantages of the former 
are discussed.   
\end{abstract}

\pacs{42.55.Ah, 42.50.Ct, 42.55.Sa, 78.67.Hc}
\maketitle
\section{Introduction}

The lasers were recognized quite early \cite{gra_hak70, 
gross_richter71,deg_scully70} as examples of systems 
undergoing a phase transition in conditions of nonequilibrium. The lasing 
regime plays the role of the 'ordered' phase, separated form the 'normal' one 
by a critical point, the laser threshold. The latter is usually identified by a 
jump in the population of the resonant cavity mode as a function of pumping. 
In a log-log plot this jump in the input-output curve is given by  
$\ln{\beta}$, \cite{rice_carm94} where the $\beta$-factor is the fraction of 
photons spontaneously emitted into the lasing mode. 

The advent of nanolasers, with few emitters and well-confined photon 
modes, allowed for $\beta$-factors close to unity, erasing the intensity jump
in the so-called 'thresholdless' lasers 
\cite{yokoyama89,khaj12,deMart88,arakawa17}. Consequently, alternative lasing 
criteria have been proposed, not always agreeing with each other.  Some are 
still based on the photon number $n$, by requiring $n>1$ 
\cite{bjork_yam94,auffeves10}. Also, coherence tests, probing the Poissonian 
photon statistics are usually applied, the most frequently invoked 
being the requirement for the second order auto-correlation function at zero 
delay, $g^{(2)}(0)$, to be close to unity \cite{cjg_lsa14, mascarenhas13, 
strauf06,ulrich07}. Other criteria have been considered as well 
\cite{cjg_lsa14}. The issue is still debated \cite{ning_what13} and has even 
prompted Nature Photonics to publish a 'checklist' \cite{checklist17} to 
provide a certain level of confidence in identifying lasing.

It is therefore surprising that a common feature of equilibrium phase 
transitions, namely the spontaneous symmetry breaking was largely ignored. 
The instability against symmetry-reducing perturbation is a characteristic of 
the ordered phase, and in this paper we address the problem of an analogue 
behavior in the lasing regime. 

We consider first the case of a single, two-level emitter interacting 
resonantly with a cavity mode. Various equivalent terminologies are 
used in the literature for the two levels, which may be seen as two atomic 
configurations, a qubit, a spin, or an electron-hole pair in a quantum dot 
which can form an exciton or recombine. 
The symmetry of this problem is the invariance with respect to an 
arbitrary common phase factor assigned to the photon and exciton quasi-spin 
operators. A coherent excitation field $E$, coupled only to the excitonic degree 
of freedom breaks this symmetry, and generates anomalous averages, i.e. 
expectation values that are strictly zero in the symmetric case. The 
'spontaneous' nature of the symmetry breaking is defined as the persistence 
of the anomalous averages in the limit of zero perturbation, as in the case of 
spontaneous magnetization in magnetic systems. 

The analysis requires that the conditions for a sharp phase transitions are 
met. Rice and Carmichael \cite{rice_carm94} have drawn the attention to the 
role of a certain limit in ensuring an abrupt transition, with a well-defined 
threshold point, in the same way as the thermodynamic limit is a necessary 
condition in the equilibrium theory. 
For the problem considered here it was shown \cite{gartner_tla11} that the 
appropriate limit is obtained by scaling down to zero both the cavity losses 
$\kappa$ and the Jaynes-Cummings (JC) coupling $g$ so that the ratio 
$g^2/\kappa$ remains finite. The order of limits is crucial, the scaling limit 
should be performed before taking $E \to 0$. We show that in this case the 
survival of the anomalous averages takes place in the parameter domain 
corresponding to the lasing regime, and only there. 

Obviously, such statements assume a proof by analytic methods. This is made 
possible by treating the system at the rate equation level, which is the 
analog of mean-field theories in equilibrium phase transitions, and which 
reduces the infinite hierarchy of equations of motion to a closed system for 
a limited set of expectation values. The treatment is a standard approach in 
quantum optics textbooks \cite{loudon00, orszag08} and, as far as the laser 
transition is concerned, it was shown to produce exact results 
\cite{gartner_tla11}. In the context of masers pumped by random injection 
\cite{scully_lamb67,stenholm73} symmetry breaking was also discussed 
in a mean-field setting \cite{deg_scully70}.

A second model addressed in the paper consists of an array of optical 
cavities each in interaction with a two-level emitter and coupled to its 
nearest neighbors by photon hopping \cite{hartmann14}. This leads to a photonic 
energy band of extended Bloch states. Applying the same procedure as above we 
show spontaneous symmetry breaking taking place in the Bloch mode resonant with 
the exciton, which is also the lasing mode. 

The paper is organized as follows: In both cases, after describing the 
models, the laser transition is first analyzed in the symmetric case, and the 
role of the scaling limit in obtaining a sharp transition is proven. Then the 
changes due to the symmetry-breaking seed are introduced. 
In the single cavity case numerical illustrations are also presented, showing 
that robustness of the anomalous averages is seen even before actually reaching 
the prescribed limits, but signaling when we are in their proximity. In this 
respect, the numerical examples show that the behavior of anomalous averages 
confirm the $g^{(2)}(0)$ criterion of lasing. This is important, since the 
former are easily accesible at the rate-equation level, while the latter is not.

\section{Single two-level emitter in a cavity}
\label{single}

In this section we analyze the case of a two-level emitter in resonance with a 
cavity mode. The emitter-photon interaction is described using the familiar 
JC Hamiltonian. Dissipative effects, like cavity losses and 
spontaneous decay of the exciton are also considered in the Lindblad 
formalism. The system is excited incoherently by an up-scattering Lindblad 
term. This problem was extensively studied in the literature, both in atomic 
\cite{agarwal90, mu_savage92} and in semiconductor quantum dot in a cavity
contexts \cite{mork08,delV_fermions09,richter_knorr09,delV_mollow10}. 

As an additional feature, we include in the Hamiltonian a symmetry-breaking 
seed in the form of an infinitesimal coherent pumping, represented by the dipole
coupling of the exciton to a resonant electric field. 

In the rotating frame the Hamiltonian reads ($\hbar=1$ throughout the paper)
\begin{equation}
H = g \, b^\dagger \sigma + g^* b \, \sigma^\dagger + 
E^*\sigma + E \sigma^\dagger\, .
\end{equation}
Here $b^\dagger, b$ are the operators of the photon mode, $\sigma^\dagger, 
\sigma$ are the pseudo-spin raising and lowering operators for the two-level 
system, $g$ is the JC coupling constant and $E$ is the strength of the coherent 
excitation. 

The equation of motion (EOM) for an arbitrary operator $A$ consists of a 
coherent, von Neuman part and of the incoherent Lindblad term contribution
\begin{align}
\frac{d}{dt} \braket{A} = &-i \braket{[A,H]} \nonumber \\
     & + \sum_{\alpha} \frac{\mu_\alpha}{2} 
     \braket{[L^\dagger_\alpha,A] L^{\phantom{\dagger}}_\alpha 
      + L^\dagger_\alpha [A, L^{\phantom{\dagger}}_\alpha]} \, ,
      \label{vNL}
\end{align}
where $\mu_\alpha$ is the rate associated with the scattering process defined 
by the operator $L_\alpha$. Three such processes are considered: the 
spontaneous relaxation of the atom (or excitonic loss in quantum dot language)
with $\mu_\alpha$ denoted by $\gamma$ and $L_\alpha=\sigma$, the cavity losses 
with the rate $\kappa$ and operator $b$, and the pumping simulated as an 
up-scattering process, $L_\alpha = \sigma^\dagger$, with the rate $P$. 

For $E=0$ the theory is $U(1)$-invariant, i.e. it is insensitive to an 
arbitrary phase factor attached to the elementary operators $e^{i \lambda}b$, 
$e^{i \lambda}\sigma$. In other words the expectation values do not depend on 
$\lambda$ and therefore non-zero averages should appear only when the operators 
are combined in phase-independent expressions, like $\braket{b^\dagger\sigma}$, 
$\braket{b^\dagger b}$ a.s.o. 
In terms of the Glauber-Sudarshan $\cal P$-representation \cite{carm_book99} 
this is related to the rotation invariance in the complex plane of the photonic 
quasi-distribution function $\cal P$. The phase symmetry is broken by the 
coherent excitation, which imposes its own phase on the system, and brings in 
anomalous averages, like 
$\braket{b}, \braket{b^\dagger}$ and $\braket{\sigma},\braket{\sigma^\dagger}$ 
As in the equilibrium phase-transition theory, by 'spontaneous' it is 
understood that anomalous averages remain nonzero even in the limit $E \to 0$. 
If and when this takes place is the subject of what follows.

Applying Eq.\eqref {vNL} one obtains the EOM for the normal averages as
\begin{subequations}
\begin{align}
\frac{d}{dt}\braket{b^\dagger b} =& -ig\braket{b^\dagger\sigma}
   +ig^*\braket{b\,\sigma^\dagger} -\kappa\braket{b^\dagger b}\, , 
                                                 \label{norm_a}\\ 
\frac{d}{dt}\braket{\sigma^\dagger\sigma} =& \,  
ig\braket{b^\dagger\sigma}
   -ig^*\braket{b\,\sigma^\dagger} -\gamma \braket{\sigma^\dagger\sigma} 
   + P\braket{\sigma \sigma^\dagger} \nonumber\\
   & +iE^*\braket{\sigma}-iE\braket{\sigma^\dagger} \, ,
                                                  \label{norm_b}\\
\frac{d}{dt}\braket{b^\dagger\sigma} =& \,
ig^*\braket{b^\dagger b\,(\sigma^\dagger\sigma-\sigma\, \sigma^\dagger)} +
ig^*\braket{\sigma^\dagger\sigma} \nonumber \\
   -&\frac{P+\gamma+\kappa}{2}\braket{b^\dagger\sigma} 
   +iE 
\braket{b^\dagger(\sigma^\dagger\sigma-\sigma\,\sigma^\dagger)} \, .
                                                  \label{norm_c}
\end{align}
\label{norm}
\end{subequations}

It is seen that in Eqs.\eqref{norm} the anomalous averages are brought in by 
the symmetry-breaking excitation. For them one derives the following EOM, 
driven by the perturbation
\begin{subequations}
\begin{align}
\frac{d}{dt}\braket{b} =& -ig \braket{\sigma}-\frac{\kappa}{2}\braket{b}
                                               \, ,       \label{anom_a}\\
\frac{d}{dt}\braket{\sigma}=&\, 
ig^*\braket{b\,(\sigma^\dagger\sigma-\sigma\, \sigma^\dagger)}
                                                    \nonumber \\
-&\frac{P+\gamma}{2}\braket{\sigma} +iE
 \braket{\sigma^\dagger\sigma-\sigma\,\sigma^\dagger} \, .
                                                    \label{anom_b}
\end{align}
\label{anom}
\end{subequations}

These equations represent the starting of an infinite hierachy of EOM. They are 
transformed into a closed system by applying a factorization approximation to 
the expectation values. The most popular is the rate equation formalism, which 
is essentially a mean-field approach to our many-body problem, and is widely 
used in the literature \cite{bjork_yam91, lorke13}. 

To be specific, in Eq.\eqref{norm_c} one factorizes level occupancies and 
photon operators in separate averages 
\begin{equation}
\braket{b^\dagger b\,(\sigma^\dagger\sigma-\sigma\, \sigma^\dagger)}
\approx\braket{b^\dagger b}(\braket{\sigma^\dagger\sigma}-\braket{\sigma\, 
\sigma^\dagger}) \, .
\label{fact_a}
\end{equation}
Similarly, in Eqs.\eqref{norm_c} and \eqref{anom_b} one uses
\begin{equation}
\braket{b\,(\sigma^\dagger\sigma-\sigma\, \sigma^\dagger)}
\approx\braket{b}(\braket{\sigma^\dagger\sigma}-\braket{\sigma\, 
\sigma^\dagger}) \, .
\label{fact_b}
\end{equation}

It is not immediately obvious why other factorizations, like $\braket{b^\dagger 
\sigma} \approx \braket{b^\dagger} \braket{\sigma}$ are not kept. In symmetric 
theories the anomalous averages involved simply do not appear, but now they are 
driven by the $E$-excitation, and in principle could contribute. 

The answer is provided by the cluster expansion theory \cite{fricke96, 
chow_jahnke13}. With averages of the elementary raising and lowering operators 
as singlets, the rate-equation limit is the systematic truncation at the 
doublet level. The following facts have to be considered: (i) Triplets and 
quadruplets are neglected. For instance, unfactorized $\braket{b^\dagger 
\sigma^\dagger \sigma}$ are not kept. (ii) In the factorization of the 
quantities appearing in the LHS of Eqs.\eqref{fact_a}, \eqref{fact_b}, all 
terms in which the operators $\sigma^\dagger$ and $\sigma$ are separated in 
different averages cancel out because of the difference involving the 
$\sigma^\dagger \sigma$ and the $\sigma \sigma^\dagger$ contributions. (iii) 
The remaining terms are of the form
\begin{equation}
\braket{b^\dagger b\,\sigma^\dagger\sigma} \approx 
\delta\braket{b^\dagger b}\braket{\sigma^\dagger \sigma} 
              + \braket{b^\dagger} \braket{b}\braket{\sigma^\dagger \sigma} \, .
              \label{clusterexp}
\end{equation}
Inserting the definition $\delta \braket{b^\dagger b}=\braket{b^\dagger 
b}-\braket{b^\dagger}\braket{b}$ one obtains the factorization shown in 
Eq.\eqref{fact_a}. A similar argument works for Eq.\eqref{fact_b}.

As a result one is left with a closed system of rate equations for the 
following unknowns: the photon number $n=\braket{b^\dagger b}$, the upper level 
(or excitonic) occupancy $f=\braket{\sigma^\dagger \sigma}$, the 
photon-assisted polarization $\psi = -ig\braket{b^\dagger \sigma}$ and the 
anomalous averages 
$\alpha = g^*\braket{b}$ and $\varphi=-i\braket{\sigma}$. The lower level 
occupancy is then $1-f=\braket{\sigma \sigma^\dagger}$. The system reads
\begin{subequations}
\begin{align}
\frac{d}{dt} n =& \, 2\psi_1 - \kappa \,n \, ,\label{rate_n}\\
\frac{d}{dt} f =&\,-2\psi_1 -\Gamma f +P -E^*\varphi-E\varphi^* 
                                           \, ,  \label{rate_f} \\
\frac{d}{dt}\psi =&\,|g|^2 n (2f-1)+|g|^2 f -\frac{\Gamma'}{2}\psi 
          + E \alpha^*(2f-1) \, ,          \label{rate_psi} \\
\frac{d}{dt}\alpha =&\,|g|^2 \varphi -\frac{\kappa}{2} \alpha  \, ,      
                                                \label{rate_alpha} \\
\frac{d}{dt}\varphi =&\, \alpha(2f-1) -\frac{\Gamma}{2} \varphi +E(2f-1)
                                                \label{rate_phi} \, .
\end{align} 
\label{rate}
\end{subequations}
Here we denoted by $\psi_1$ the real part of $\psi$, $\Gamma = P + \gamma$ and
$\Gamma'=P+\gamma+\kappa$. 
Before analyzing the solution in the limit $E \to 0$, we summarize below 
the $E=0$ situation.

\subsection{The laser transition in the scaling limit}
\label{no_E}

In the absence of the symmetry-breaking term one is left with the first three 
Eqs.\eqref{rate}, for the unknowns $n$, $f$ and $\psi$. In the steady state the 
time derivatives are vanishing and the system becomes algebraic. One notices 
that $\psi$ becomes real, $\psi_1=\psi$, and is given by Eq.\eqref{rate_psi} as
\begin{equation}
2\psi = R'n(2f-1)+R'f \, , \label{psi}
\end{equation}
with $R'=4|g|^2/\Gamma'$. We use Eq.\eqref{rate_n} to eliminate $\psi$
via $2\psi=\kappa n$, and stay with the main variables, the photon and the 
upper-level populations.
\begin{subequations}
\begin{align}
\kappa n = & R'n (2f-1)+R'f \, , \label{balance_a} \\
\kappa n = & -\Gamma f +P \, . \label{balance_b}
\end{align}
\label{balance}
\end{subequations}

The first of these equation is simply the photon balance between the loss 
through the cavity walls and the net generation, consisting of spontaneous and 
stimulated emission $R'(n+1)f$ minus absorption $R'n(1-f)$, as in the theory of 
Einstein coefficients \cite{loudon00}, with $R'$ identified as the 
spontaneous emission rate. The condition can be recast in the form 
\begin{equation}
f_L'-f = \frac{f}{2n} \qquad \mathrm{with}\qquad 
f_L'=\frac{1}{2} + \frac{\kappa}{2R'} \,.\label{f_Lprime}
\end{equation}
It is easy to see that $f_L'$ is the population for which the gain in the 
active medium (stimulated emission minus absorption) exactly compansated the 
cavity losses. One consequence of Eq.\eqref{f_Lprime} is that $f_L'$ is an 
upper bound for the solution, $f \leqslant f_L'$ for all pump values. 

The second equation is again a balance condition, expressing the fact that in 
the steady state the loss of an excitation, either photonic $\kappa n$, or 
excitonic $\gamma f$, should be compensated by the pumping process $P(1-f)$. 
The condition can be rewritten as
\begin{equation}
f_N-f = \frac{\kappa n}{\Gamma }\qquad \mathrm{with}\qquad 
f_N=\frac{P}{P+\gamma} \, , \label{f_N}
\end{equation}
where $f_N$ represents the excitonic occupancy of the isolated emitter ($g=0$). 
Again one obtains an upper bound $f \leqslant f_N$ for all $P$. 

\begin{figure}
  \centering
   \includegraphics[width=0.45\textwidth]{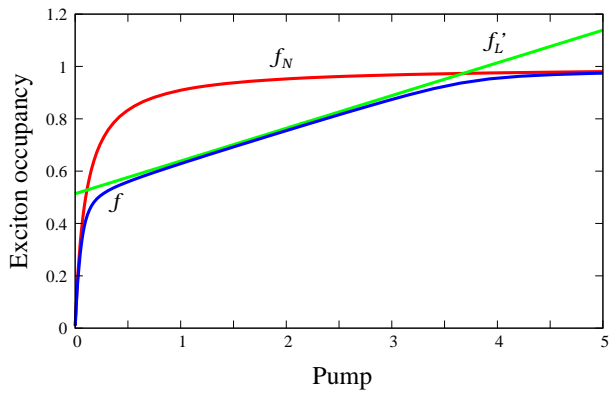}
   \caption{The limit solutions $f_N$ (red) and $f_L'$ (green)
   are upper bounds for the true one (blue). The parameters used are
   $\gamma = 0.1$, $\kappa = 0.01$ and $g = 0.1$. 
   \label{fig:Fig1}}
\end{figure}

By multiplying the Eqs.\eqref{f_Lprime} and \eqref{f_N} one obtains a 
quadratic equation for the exciton population
\begin{equation}
(f_N-f)(f_L'-f)=\frac{\kappa}{2(P+\gamma)}\,f \, , \label{eff}
\end{equation}
whose lower solution $f < \min\{f_N, f_L'\}$ is the physical one. 
One notices that for small values of $\kappa$ the RHS of this equation becomes 
small, and this pushes the solution close to either one of the limit cases, 
$f_N$, or the other, $f_L'$, whichever is lower (see Fig.\ref{fig:Fig1}).
Therefore, with increasing $P$, the system changes between two regimes, 
characterized by either $f \approx f_N$ or $f \approx f_L'$. As long as the RHS 
of Eq.\eqref{eff} is nonzero the transition is smooth, with no well-defined 
threshold. 

In order to obtain an abrupt transition, with precise transition points, one 
has to perform some limiting procedure. The obvious limit is $\kappa \to 0$, 
which brings the RHS of the Eq.\eqref{eff} to zero, and thus leads the exciton 
population as function of pumping to develop angular points, as it shifts from 
one limit solution to the other. On the other hand, in order to preserve the 
qualitative picture of Fig.\ref{fig:Fig1}, one simultaneously takes the limit 
$g \to 0$ so that $g^2/\kappa$ remains finite. This amounts to only a minor 
change in $f_L'$, which becomes now 
\begin{equation}
f_L' \to f_L = \frac{1}{2}+\frac{\kappa}{2R} = \frac{1}{2}+
    \frac{\kappa (P+\gamma)}{8|g|^2} \, ,
\end{equation}
where the limit of $\kappa/R'$ is $\kappa/R$, with $R=4|g|^2/\Gamma= 
4|g|^2/(P+\gamma)$. Without 
the $g \to 0$ limit one would have the trivial result $f_L' \to 1/2$. It is 
worth noting that the $\beta$-factor is proportional to $|g|^2$, and 
therefore this can be also viewed as a small $\beta$ limit, as required for a 
sharp transition \cite{rice_carm94}.

The conclusion is that in the limit in which both $g$ and $\kappa$ are scaled 
down to zero, while keeping the proportion $g^2/\kappa$ finite, one has a sharp 
transition between $f=f_N$ and $f=f_L$. The non-analyticity of the solution is 
represented by angular points appearing at the $f_N=f_L$ crossing points. 

Turning now to the photon population, we have from Eq.\eqref{f_N}
\begin{equation}
\kappa n = (P+\gamma)(f_N-f)\, . \label{enn}
\end{equation}
Again, two contrasting situations appear. In the $f=f_N$ regime it is clear 
that the value of photon number cannot prevent the product $\kappa n$ to vanish 
in the scaling limit. This is the normal, non-lasing phase (hence the index 
N). On the contrary, in the case $f=f_L$ the photon number 
goes to infinity, such that the product $\kappa n$ stays finite. This 
is indicative of "an explosion of stimulated emission" \cite{rice_carm94}, 
and identifies the lasing regime (hence the index L). 

Note that the condition for the existence of real roots $P$ for the quadratic
equation $f_N=f_L$ is $|g|^2/\kappa > 2 \gamma$ \cite{gartner_tla11}. This is 
the lasing condition upon the parameters. 
When it is met there are two crossing points. Going from low 
to high pump values the system is initially normal, up to the first crossover. 
This is the threshold value $P_{th}$, where the system starts lasing. 
Reaching the second crossing it becomes normal again, due to the so-called 
quenching phenomenon \cite{mu_savage92}, which is a consequence of the 
excitation-induced dephasing.   

In the absence of a size parameter one cannot define the thermodynamic limit 
for our problem, as in the theory of equilibrium phase transition. This role is 
played here by the scaling limit. In this case a "macroscopic" photon 
population corresponds to one that grows like $1/\kappa$. It might seem trivial 
that $n$ increases when the cavity quality is getting better, but one should 
keep in mind that in the scaling limit the rate of photon generation, $R \sim 
|g|^2$, is reduced too, and precisely in the same ratio as the loss rate. 

Before closing this discussion we note that the proof of these results can be 
made rigorous, not depending on the rate-equation approximation 
\cite{gartner_tla11}.   

\subsection{Spontaneous symmetry breaking for the two-level laser}
\label{with_E}

In the presence of the symmetry-breaking excitation, Eqs.\eqref{rate_alpha} and 
\eqref{rate_phi} for the anomalous averages are driven by the $E$-term. 
Solving for steady-state values of $\alpha$ and $\varphi$ gives
\begin{subequations}
\begin{align}
\alpha = & E \,\frac{f-\frac{1}{2}}{f_L-f} \, ,  \label{anml_alpha} \\
\varphi =& E\, \frac{\kappa}{2|g|^2} \, \frac{f-\frac{1}{2}}{f_L-f} \, . 
\label{anml_phi}
\end{align}
\label{anml}
\end{subequations}
The anomalous averages are driven by $E$, their phases are the phase 
of $E$ and can be easily factored out. Therefore, and for the sake of 
simplicity, we will assume that $E$ is real and positive. 

It is already clear that these averages may survive in the limit $E \to 
0$ only if simultaneously $f \to f_L$, in a way that keeps the ratio at a 
finite value. We have seen that in the scaling limit, and in the lasing regime, 
$f$ does approach $f_L$. It remains to analyze the interplay of the two 
limit procedures. 

Introducing these results in Eqs.\eqref{rate_n}-\eqref{rate_psi}, one notices 
again that $\psi$ is real, obeying now the relation
\begin{equation}
2\psi = \kappa n = R'n(2f-1)+R'f + \frac{8E^2}{\Gamma'} 
\,\frac{(f-\frac{1}{2})^2}{f_L-f}\, , \label{psi_new}
\end{equation}
which is the extension of Eq.\eqref{balance_a}, and can be recast as
\begin{equation}
\kappa n(f_L'-f)=\frac{\kappa}{2}f +E^2 \frac{\kappa}{\,|g|^2}\, 
\frac{(f-\frac{1}{2})^2}{f_L-f} \, . \label{bal_lam_a}
\end{equation}
The generalization of Eq.\eqref{balance_b} reads
\begin{equation}
\kappa n =  (P+\gamma)(f_N-f)
-E^2\,\frac{\kappa}{\,|g|^2}\,\frac{f-\frac{1}{2}}{f_L-f} \, .
 \label{bal_lam_b}
\end{equation}
Now it is easy to eliminate $n$ and one obtains for $f$ the cubic equation
\begin{align}
\Big[ (f_N-f)(f_L'-&f)-\frac{\kappa}{2(P+\gamma)}f \Big] \,(f_L-f) 
                                                            \nonumber \\
  - & \,E^2\, \frac{\kappa^2 \, \Gamma'}{8 \,|g|^4\, \Gamma} \, 
      \,(f-1/2) = 0 \, , \label{cubic}
\end{align}
where the quadratic polynomial in the square brackets, denoted in what follows 
by $Q(f)$, provides the roots for the symmetric problem, see Eq.\eqref{eff}. 

The natural question arising in the presence of two limit procedures concerns 
their order. We will show below that performing first the limit $E \to 0$,
the anomalous averages go to zero and one recovers the symmetric case results 
of Section \ref{no_E}.  Applying subsequently the scaling limit one reaches a 
sharp laser transition, as described there. This is the expected behavior, but 
the proof requires some attention. On the contrary, if the scaling limit is 
performed first, the anomalous averages remain non-zero even after the $E 
\to 0$ limit, and this spontaneous symmetry breaking takes place in the 
lasing regime. Below we analyze the two limit orders.

(i) {\em If the $E \to 0$ limit is performed first}, Eq.\eqref{cubic} 
becomes $Q(f)\,(f_L-f)=0$. We analyze first possibility is that its solution is 
the lower root of $Q(f)=0$, as in the symmetric case. We know that this root 
obeys $f< f_N, f_L'$, but in order to prove that the anomalous averages 
disappear indeed, one has to show that $f$ stays away from $f_L$. This is not 
immediate, since $f_L$ too, like $f$, is on the lower side of $f_L'$. We prove 
that actually $f$ is strictly smaller than $f_L$ by checking that $Q(f_L)$ is 
negative. Indeed, using the explicit expressions of $f_N, f_L'$ and $f_L$ one 
obtains
\begin{align}
Q(f_L) = & \left(\frac{P-\gamma}{2(P+\gamma)}-\frac{\kappa (P+\gamma)}{8|g|^2} 
       \right)\,\frac{\kappa^2}{8|g|^2}
       \nonumber \\
       - & \frac{\kappa}{2(P+\gamma)}\, 
        \left(\frac{1}{2}+\frac{\kappa(P+\gamma)}{8|g|^2} \right)\, .
\end{align}
Here one notices that the only positive contribution comes from the first term, 
proportional to $P$, which is exactly cancelled by the last $P$-term. All 
the other terms being negative one has $Q(f)<0$, which proves that $f_L$ is 
placed between the roots, i.e. $f<f_L$. 

A second possibility to be considered is that, in the limit $E \to 0$, 
the physical solution converges to $f_L$. This can be ruled out by noticing 
from Eq.\eqref{cubic} that in such a case $f_L-f \sim E^2$ and then the 
anomalous average $\varphi$ would go to infinity. But this is impossible, 
since $\varphi$ is the expectation value of $\sigma$, a bounded operator. 

(ii) {\em If the scaling limit is performed first}, Eq.\eqref{cubic} becomes
\begin{equation}
 (f_N-f)(f_L-f)^2\, - \,E^2\, 
   \frac{\kappa^2}{8 \,|g|^4} \,(f-1/2) = 0 \, , 
    \label{scale_cubic}
\end{equation}
After that, in the $E \to 0$ limit, one possibility is $f \to 
f_N$, which is the normal phase. But in the lasing regime we see that
$f$ converges to $f_L$ in such a way that $f_L-f \sim E$, which is exactly the 
order in the infinitesimal parameter that keeps the anomalous averages 
non-zero. 

This statement is the main result of the paper: {\em spontaneous symmetry 
breaking does take place, but only in the lasing phase as it is defined by the 
scaling limit.} Additionally, the result stresses that the scaling limit is 
instrumental in understanding the laser transition.  

The photon population behavior in the scaling limit is contained in  
Eq.\eqref{bal_lam_a}, which has now the form
\begin{equation}
\kappa n = E^2 \frac{\kappa}{\,|g|^2}\, 
\frac{(f-\frac{1}{2})^2}{(f_L-f)^2}=\frac{\kappa}{\,|g|^2}\, \alpha^2 \, ,
\end{equation}
which amounts to $\braket{b^\dagger b}=|\langle b \rangle|^2$. This is what is 
heuristically expected from a coherent photonic state, in which the 
$b$ operator behaves like a c-number.  

\begin{figure}
  \centering
   \includegraphics[width=0.4\textwidth]{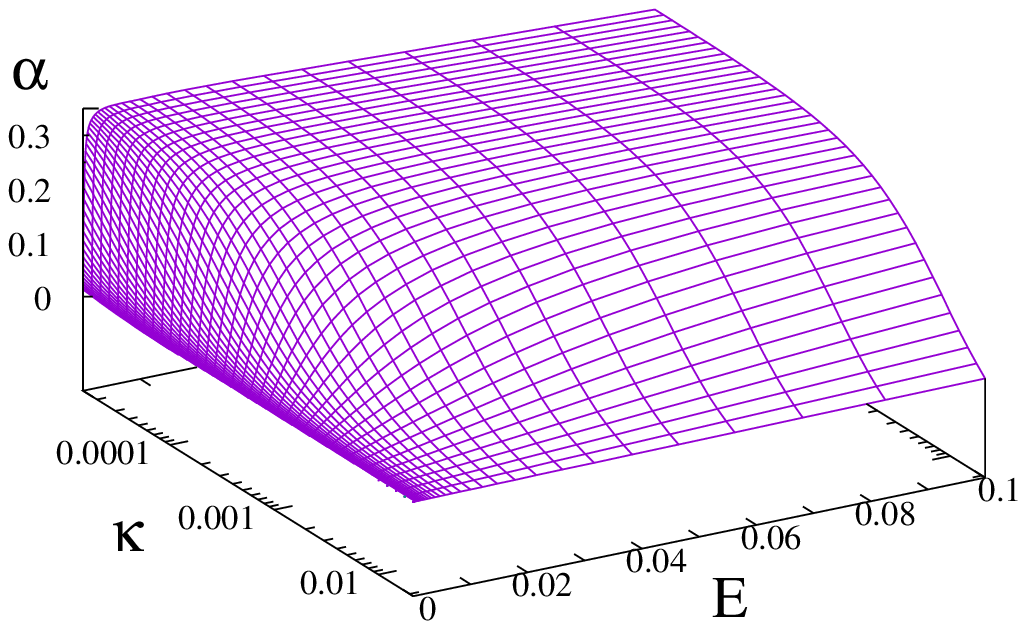}
   
   \vspace{0.3cm}
   \includegraphics[width=0.4\textwidth]{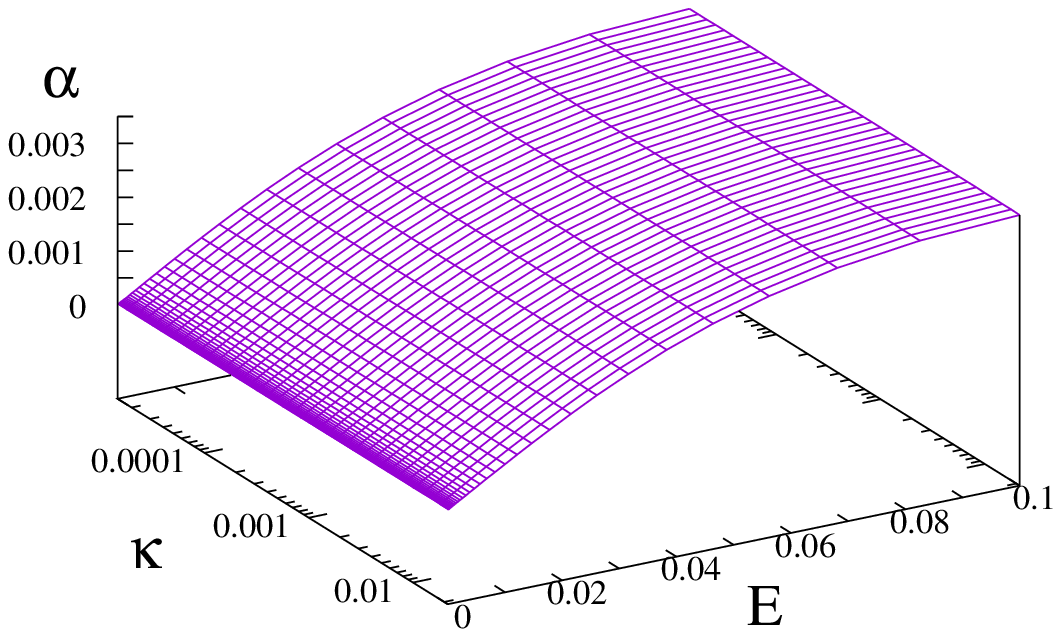}   
   \caption{The anomalous average $\alpha$ as function of the 
symmetry-breaking seed $E$, for different values of $\kappa$.
Other parameters: $g^2=\kappa$, $P=0.3$, $\gamma=0.02$ 
(upper panel) and $g^2=\kappa/200$, $P=0.3$, $\gamma=0.01$ (lower panel).  
   \label{fig:Fig2}}
\end{figure}

From a practical point of view the limits discussed above cannot be reached 
numerically, and even less experimentally. Nevertheless, one may bring the 
parameters in the asymptotic domain sufficiently close to the limits to see 
their influence, in the sense of predicting with good accuracy the behavior of 
the system. The correct limit order is simulated by taking $\kappa, g^2 \ll 
E$. 

Indeed, as seen in Fig.\ref{fig:Fig2}, upper panel, the anomalous average 
$\alpha$ goes to zero with $E$, but the convergence gets 
significantly delayed by decreasing $\kappa$ and $g^2$, as a numerical 
hint that the scaling limit prevents the vanishing of $\alpha$ altogether.
This is the situation for parameters corresponding to the laser regime. On the 
contrary, in the lower panel the parameters do not meet the lasing condition 
and one sees that the decrease of $\alpha$ with $E$ is completely insensitive 
to the $\kappa$ values. Note also the absolute values, which are two orders of 
magnitude lower.

\begin{figure}
  \centering
   \includegraphics[width=0.4\textwidth]{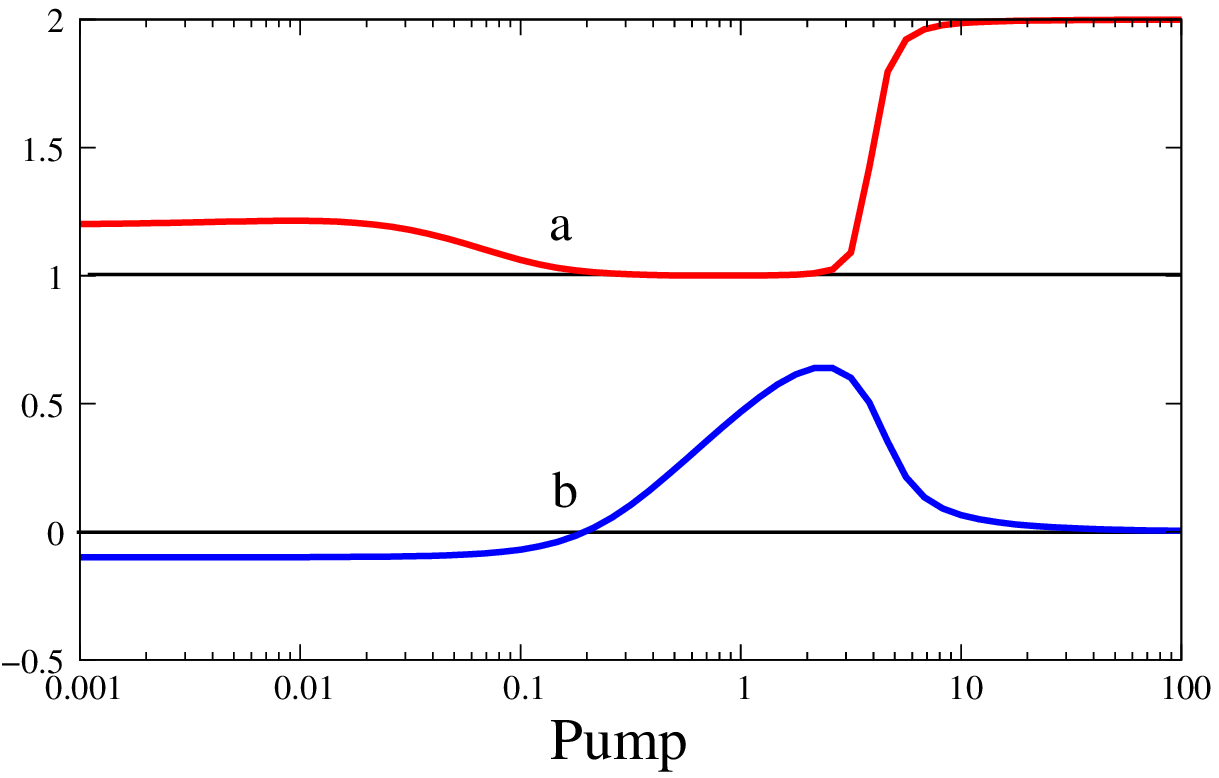}
   \includegraphics[width=0.4\textwidth]{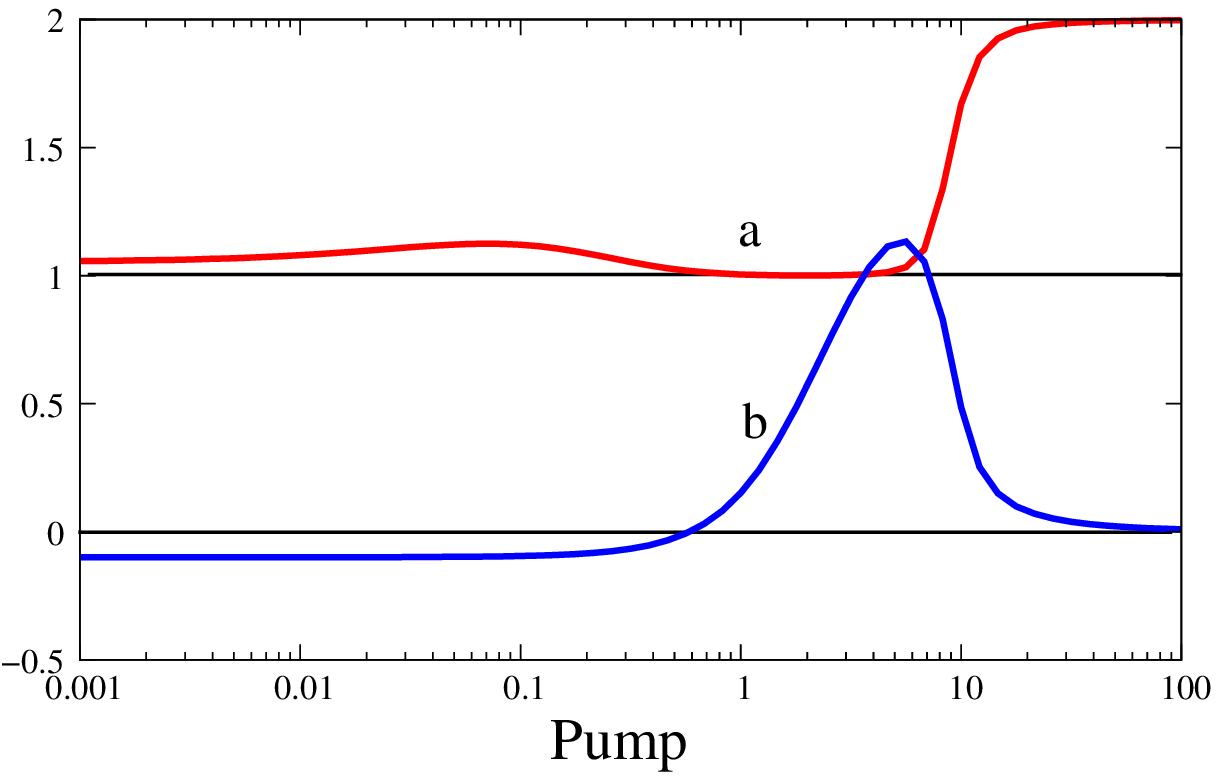}   
   \caption{ (a) $g^{(2)}(0)$ and (b) $\alpha$, as functions 
of pumping. The parameters are: $\kappa=0.01$ , $g=0.1$ , 
$\gamma=0.02$ for the upper panel and $\kappa=0.04$ , $g=0.3$ , 
$\gamma=0.05$ for the lower one. In both $E=0.1$. Horizontal lines at 1 
and 0 are guide to the eye.
   \label{fig:Fig3}}
\end{figure}

It is also worth noting that $\alpha$ shows a positive bump in the pumping 
interval in which the photon autocorrelation function at zero time 
delay, $g^{(2)}(0)$, is close to unity. The latter is currently used as a test 
of coherent light, but unfortunately it is not accessible at the rate equation 
level, being related to the quadruplet $\braket{b^\dagger b^\dagger b\,b}$. 
For instance, in the cases illustrated in Fig.\ref{fig:Fig3}, in order to 
calculate $g^{(2)}(0)$ we had to solve the full, infinite hierarchy by using a 
continued fraction method \cite{agarwal90, gartner_tla11}. In contrast the 
anomalous average $\alpha \sim \braket{b}$ is a singlet, and thus it was easy 
to obtain. 

The situations shown in Fig.\ref{fig:Fig3} are relatively far from $E=0$, and 
in the lower panel further away from the scaling limit too. Yet both cases 
show that the examination of $\alpha$ is a useful tool in its own right for 
signaling the lasing behavior. One should also keep in mind that $g^{(2)}(0)=1$ 
is a necessary but not sufficient condition for coherent light.


\section{Cavity arrays}
\label{chain}

In this section we show that the analysis above and its conclusions are not 
limited to the simple case of a single, two-level emitter. Here we consider the 
more complicated system of a linear array of cavities, coupled by a photon 
hopping term, describing the possible leaking from one cavity to its neigbors. 
A two-level emitter is placed in each of these cavities and is coupled to its 
photon mode by the JC interaction.
  
The Hamiltonian of such a system is 
\begin{align}
H =\, & \omega \sum_r b^\dagger_r b^{\phantom{\dagger}}_r 
   +J\sum_r \left [b^\dagger_r b^{\phantom{\dagger}}_{r+1} 
   +b^\dagger_{r+1} b^{\phantom{\dagger}}_r\right] \nonumber \\ 
   +\,& \varepsilon \sum_r \sigma^\dagger_r \sigma^{\phantom{\dagger}}_r 
   + \sum_r \left [g \, b^\dagger_r \sigma^{\phantom{\dagger}}_r 
   + g^* b^{\phantom{\dagger}}_r\,\sigma^\dagger_r \right ] \, . 
\end{align}
The chain of $N$ cavities is assumed homogeneous, with the photon frequency 
$\omega$ and the exciton energy $\varepsilon$ independent on the position $r$ 
on the chain.  The same holds for the JC coupling $g$ and for the hopping term 
$J$ to nearest neigbors. Assuming cyclic boundary conditions and taking 
advantage of the translation invariance, the photon part of the Hamiltonian
becomes diagonal in the plane-wave representation
\begin{equation}
H_{ph}= \sum_k \omega_k b^\dagger_k b^{\phantom{\dagger}}_k \, , 
\end{equation}
where $\omega_k = \omega +2J \cos k$ are the Bloch-mode frequencies, forming 
the energy band of photonic eigenstates. The momentum $k$ runs on the 
Brillouin zone (BZ), represented by the interval $(-\pi, \pi]$.
The corresponding operators are connected to the local ones by 
\begin{equation}
b_k = \frac{1}{\sqrt{N}}\sum_r e^{ikr} b_r \, ,\; \;\text{and} \;\;
b_r = \frac{1}{\sqrt{N}}\sum_k e^{-ikr} b_k \, .
\label{bloch_b}
\end{equation}
To the above Hamiltonian a symmetry-breaking seed is added in the form
\begin{equation}
H_{sb}=\sum_r \left [ E^*_r(t) \sigma^{\phantom{\dagger}}_r 
       +E^{\phantom{\dagger}}_r(t) \sigma^\dagger_r \right] \, ,
\label{sb}
\end{equation}
where $E_r(t) = E(t) e^{iqr}$, is chosen to excite a particular Bloch mode 
$q$. The translation invariance is spoiled by this $r$-dependence, but it is 
formally recovered by the unitary transform
\begin{equation}
\sigma_r \to e^{iqr}\sigma_r \, , \;\; b_r \to e^{iqr}b_r \,
\end{equation}
which restores the expression of the translation invariant Hamiltonian up to a 
shift of the photon spectrum
\begin{equation}
H_{ph}=\sum_k \omega^{\phantom{\dagger}}_{k-q}b^\dagger_k 
    b^{\phantom{\dagger}}_k \, .
\end{equation}
In the rotating frame with respect to 
\begin{equation}
H_0 = \varepsilon \sum_k b^\dagger_k b^{\phantom{\dagger}}_k 
    + \varepsilon \sum_r \sigma^\dagger_r \sigma^{\phantom{\dagger}}_r \, ,
\end{equation}
and considering a resonant coherent excitation $E(t) = E e^{-i\varepsilon t}$,
one is left with the Hamiltonian
\begin{align}
H=& - \sum_k \Delta^{\phantom{\dagger}}_{k-q} b^\dagger_k 
b^{\phantom{\dagger}}_k 
    + \sum_{r,k} \left [g_r(k) \, b^\dagger_k \sigma^{\phantom{\dagger}}_r + 
g^*_r(k) b_k\,\sigma^\dagger_r \right ] \nonumber \\
  & + \sum_r \left[ E^* \sigma_r + E \sigma^\dagger_r \right ] \, ,   
\end{align} 
where $\Delta_k = \varepsilon- \omega_k$ is the $k$-mode detuning and $g_r(k) = 
g\, e^{ikr}/\sqrt{N}$. The $k,r$ double summation could be avoided by defining
$\sigma_k$ as the Fourier transform of $\sigma_r$, but this is not as useful 
for what follows as Eq. \eqref{bloch_b} because of the more complicated  
commutation relations ensuing. 

The dissipative part of the model consists of including at each position $r$ on 
the chain Lindblad terms identical to those described in Eq.\eqref{vNL}. For 
instance, for the cavity loss  one has
\begin{equation}
\frac{d}{dt} \braket{A}_\text{cav.loss}=\frac{\kappa}{2}\sum_r 
\braket{[b^\dagger_r,A] b^{\phantom{\dagger}}_r 
    + b^\dagger_r [A,b^{\phantom{\dagger}}_r]} \, ,
\end{equation}
which, by unitarity, has the same expression in the $k$-representation.

The EOM for this system, considered at the rate equation level involves a 
limited set of expectation values. We have the photon Bloch-mode populations, 
which is the Fourier transform of the photon-photon correlation along the chain
\begin{equation}
n_k= \braket{b^\dagger_k b^{\phantom{\dagger}}_k} = \sum_r 
e^{-ikr}\braket{b^\dagger_r b^{\phantom{\dagger}}_0} \, .
\end{equation}
We made use here of the translation invariance property $\braket{b^\dagger_r 
b^{\phantom{\dagger}}_{r'}}= \braket{b^\dagger_{r-r'}b^{\phantom{\dagger}}_0}$. 
The inverse relation reads
\begin{equation}
\braket{b^\dagger_r b^{\phantom{\dagger}}_0}=\frac{1}{N}\sum_k e^{ikr} n_k \, ,
\end{equation}
from which one obtains the on-site population as a BZ average 
of the mode population 
\begin{equation}
n = \braket{b^\dagger_0 b^{\phantom{\dagger}}_0}=\frac{1}{N}\sum_k n_k \, .
\end{equation}
Again, by translation invariance, the on-site population is the same on all 
sites, and therefore its notation does not carry an index.

Similar quantities arise in connection with the excitonic populations and 
correlations
\begin{align}
& f_k = \braket{\sigma^\dagger_k \sigma^{\phantom{\dagger}}_k}  
    =\sum_r e^{-ikr}\braket{\sigma^\dagger_r \sigma^{\phantom{\dagger}}_0} \, \\
& \braket{\sigma^\dagger_r \sigma^{\phantom{\dagger}}_0} 
    = \frac{1}{N}\sum_k e^{ikr} f_k \, ,
\end{align}
and the on-site exciton population 
$f=\braket{\sigma^\dagger_0 \sigma^{\phantom{\dagger}}_0}$ is 
obtained as the BZ average of $f_k$. It is obvious from the definitions that 
$n_k$ and $f_k$ are real, positive quantities. In a previous paper 
\cite{hartmann14} the $\sigma$-$\sigma$ correlators were neglected, so that in 
this respect the present treatment is slightly more general. 

One also encounters mixed, photon-exciton correlators, and it is convenient to 
define, in analogy with the single cavity case
\begin{equation}
\psi_k = -ig \sqrt{N} \braket{b^\dagger_k \sigma^{\phantom{\dagger}}_0} \, .
\end{equation}
The site $r=0$ does not play a special role since, by translation invariance 
one has 
$\braket{b^\dagger_k \sigma^{\phantom{\dagger}}_r} =e^{-ikr} \braket{b^\dagger_k 
\sigma^{\phantom{\dagger}}_0}$. 

The anomalous averages one has to consider are
\begin{align}
\alpha_k = \, & g^* \sqrt{N} \braket{b_k} 
        = g^*\braket{b_r} N \delta_{k,0}=\alpha N \delta_{k,0} \, ,\\
\varphi_k =&-i \sqrt{N} \braket{\sigma_k} 
          = -i \braket{\sigma_r} N \delta_{k,0} = \varphi N \delta_{k,0} \, .
\end{align}
The $r$-dependence is spurious, the sites being identical. This brings the 
$k$-depenence to a $\delta$-function located at $k=0$. As before, the 
indexless $\alpha, \varphi$ denote the BZ averages of the respective 
$k$-dependent quantities. Note that $N \delta_{k,0}$ becomes $2 \pi \delta(k)$ 
in the infinite chain limit. 

With these notations the EOM have the form 
\begin{subequations}
\begin{align}
\frac{d}{dt} n_k =\, & 2 \psi_{k,1} - \kappa n_k \, , \label{ch_norm_a}\\
\frac{d}{dt} f_k =\, & P-\Gamma f_k + 2\psi_{k,1}(2f-1) - 2\psi_1 \, 2f
   \nonumber \\
    + & (E^*\varphi_k + c.c.)(2f-1) -(E^* \varphi+c.c.) 2f \, 
    \label{ch_norm_b}\\
\frac{d}{dt} \psi_k =\, & -\left (\frac{\Gamma'}{2} +i 
     \Delta_{k-q}\right )\psi_k + |g|^2 n_k (2f-1) \nonumber \\ 
     & + |g|^2 f_k +E \alpha^*_k (2f-1) \, . \label{ch_norm_c} 
\end{align}
\label{ch_norm}
\end{subequations}
The subindex $1$ in $\psi_{k,1}$ and $\psi_1$ denotes the real part, the 
absence of the momentum index implies the BZ average, and $c.c.$ is short for 
complex conjugate. 
Similitudes and differences to Eqs.\eqref{norm} are obvious. The nonlinear 
terms stem from factorizations of the same kind as used there, but slightly 
more complicated. For instance we split populations from photon-assisted 
polarizations in separate factors, i.e.  
\begin{equation}
\braket{b^{\phantom{\dagger}}_k \sigma^\dagger_r 
[\sigma^{\phantom{\dagger}}_0,\sigma^\dagger_0]} 
     \approx \braket{b^{\phantom{\dagger}}_k\sigma^\dagger_r}
     \braket{[\sigma^{\phantom{\dagger}}_0,\sigma^\dagger_0]}
     = e^{ikr}\braket{b^{\phantom{\dagger}}_k \sigma^\dagger_0}(1-2f)\, ,
\end{equation}
for $r \neq 0$, but one has exactly 
$\braket{b^{\phantom{\dagger}}_k \sigma^\dagger_r 
[\sigma^{\phantom{\dagger}}_0,\sigma^\dagger_0]}
=\braket{b^{\phantom{\dagger}}_k \sigma^\dagger_0}$ if $r=0$. 
These single-site terms lead to the $k$-independent subtractions appearing in 
Eq.\eqref{ch_norm_b}. The EOM for the anomalous averages read
\begin{subequations}
\begin{align}
\frac{d}{dt} \alpha_k =\, &-\left(\frac{\kappa}{2}-i \Delta_{k-q}\right) 
             \alpha_k + |g|^2 \varphi_k \, , \label{ch_anom_a} \\
\frac{d}{dt} \varphi_k =\, &-\frac{\Gamma}{2}\varphi_k + \alpha_k (2f-1) 
                       +E(2f-1)\, N \delta_{k,0} \, . \label{ch_anom_b}    
\end{align}
\label{ch_anom}
\end{subequations}

\subsection{Laser transition in cavity arrays}
\label{ch_no_E}

As in the single cavity case, we analyze first the laser transition in the 
absence of symmetry breaking. To this end we consider the steady-state solution 
of Eqs.\eqref{ch_norm} for $E=0$. It is easy to 
eliminate $\psi_{k,1}$ using the first equation: $2 \psi_{k,1}= \kappa n_k$. As 
a consequence one has also $2 \psi_1= \kappa n$. From Eq.\eqref{ch_norm_b} one 
obtains succesively
\begin{equation}
f_k= f_N + (2f-1)\frac{\kappa}{P+\gamma} n_k  - 2f \frac{\kappa}{P+\gamma} n \,,
\label{f_k}
\end{equation}
then, after averaging over the BZ 
\begin{equation}
\frac{\kappa}{P+\gamma} n = f_N-f \, ,
\label{ch_f_N}
\end{equation}
which is used to rewrite Eq.\eqref{f_k} as
\begin{equation}
f_k= (2f-1)\frac{\kappa}{P+\gamma} n_k + \left[ f_N - 2f (f_N-f) \right] \, .
\label{f_k_Phi}
\end{equation}
This is a first equation connecting the populations $f_k$ and $n_k$. 
The term in square brackets is quadratic in $f$ and in what follows will be 
denoted by $\Phi(f)$ or just $\Phi$. It is important to note that $\Phi(f)$
remains a positive quantity for any $f$.

The relation \eqref{ch_f_N} is the analogue of Eq.\eqref{f_N} and translates 
the fact that the Hamiltonian conserves the total excitation number $\sum_r 
\left[b^\dagger_r b_r + \sigma^\dagger_r \sigma_r \right]$, which is influenced 
only by the dissipation terms. Also, an important consequence is that 
one recovers the same upper bound $f \leqslant f_N$.   

A second equation is derived from Eq.\eqref{ch_norm_c} and reads
\begin{equation}
\kappa n_k = (2f-1) R'_k n_k + R'_k f_k \, ,
\label{n_k}
\end{equation}
where
$R'_k = 4|g|^2 \Gamma'/(\Gamma'^2 + 4 \Delta^2_k)$ 
denotes the spontaneous transition rate into mode $k$, depending on its 
detuning $\Delta_k$. This is the counterpart of Eq.\eqref{balance_a}.
Now one can eliminate $f_k$ from the two equations and obtain $n_k$
as a function of $f$ only. The result can be put in the form
\begin{equation}
n_k = \frac{\frac{|g|^2}{\kappa} \Gamma' \Phi(f)}{\Delta^2_k+ 
2\frac{|g|^2}{\kappa}\frac{\Gamma'^2}{\Gamma}(f_L-f)} \, ,
\label{ennka}
\end{equation}
in which, in the denominator, the detuning term was separated. Then, by 
averaging the result over the BZ and using again Eq.\eqref{ch_f_N}, we obtain a 
closed equation for $f$. In the limit of large $N$ the BZ averaging is 
expressed by a Lorentzian integral with respect to the detuning 
\begin{equation}
f_N-f = |g|^2 \frac{1}{2 \pi}\int_{-\pi}^{\pi} \frac{A}{(2\cos k-\Delta)^2+ 
         \delta^2}\, d k \, .
\label{lorentz}         
\end{equation}
Here $\Delta= (\varepsilon-\omega)/J$ is the on-site detuning normalized to the 
hopping rate $J$ and $A= (\Gamma' \Phi)/(J^2 \Gamma)$. Also
\begin{equation}
\delta = \frac{\Gamma'}{J}\sqrt{\frac{2|g|^2}{\kappa \Gamma}} \sqrt{f_L-f}=B 
         \sqrt{f_L-f} \,.
\end{equation}
The notation used for the  $\delta^2$ term in the denominator 
of Eq.\eqref{lorentz} suggests that the quantity is positive,
which is by no means obvious. 
Nevertheless, the positivity of the Bloch-mode populations $n_k$ 
entails the positivity of the denominator in Eq.\eqref{ennka} for all $k$. If 
additionally we assume that there are resonant modes in the system, $\Delta_k 
=0$ for some $k$, then $\delta^2$ must be positive indeed. This is the case we 
consider from now on. Then $\delta$ is real and moreover, by this argument we 
also recover the second upper bound $f\leqslant f_L$. 

Eq.\eqref{lorentz} can be solved only numerically, since the unknown function 
$f$ is expressed as an integral involving other $f$-dependent quantities, like 
$\Phi$ and $\delta$. Still, the behavior of the solution in the scaling limit 
is analytically accessible. The constants $A$ and $B$ are stable (neither 
vanishing, nor divergent) in this limit since $g$ and $\kappa$ appear in the 
scaling ratio $g^2/\kappa$ and $\Gamma' \to \Gamma$. Only the prefactor $|g|^2$
is not compensated, and seems to imply that one has necessarily $f \to f_N$. 
But this contradicts the inequality $f \leqslant f_L$, when $f_L < f_N$.

The solution of the paradox relies on the possibility that the integral itself 
becomes divergent, as indeed is the case, due to the existence of resonant 
modes, and provided $\delta$ also vanishes, i.e. $f \to f_L$ in the scaling 
limit.  
 
This heuristic argument can be made precise by performing the integral 
analytically \cite{hartmann14}, which is done by mapping the $[-\pi, \pi]$ 
interval over the unit cercle in the complex plane and using the residue 
theorem. The result is
\begin{equation}
(f_N-f) \sqrt{f_L-f} = |g|^2 \frac{A}{B} \,\Re \left 
\{\frac{i}{\zeta_2-\zeta_1}\right \} \, ,
\label{ch_eff}
\end{equation}
where $\zeta_{1,2}$ are the roots of the quadratic equation $z^2-uz+1=0$, with
$u= \Delta +i\delta$, obeying $|\zeta_1|<1<|\zeta_2|$. 

The conclusion is again that in the scaling limit $f$ converges to either $f_N$ 
or $f_L$, whichever is the smaller. As the pumping increases, the transition 
between these options is sharp, as in the single emitter case. The macroscopic 
photon population occurs only in the $f=f_L$ regime. It is now obvious that the 
macroscopically occupied modes are the resonant ones.  

As a final remark, we note that the obvious main difference from Eq.\eqref{eff} 
appears in the reduced power of $f_L-f$. This is a consequence of the 
integration over $k$, which reduces the singularity $1/\delta^2$ of the 
integrand to $1/\delta$.
The factor $i/(\zeta_2-\zeta_1)$, related to the photonic density of states, 
introduces additional singularities only if the resonant mode is at the 
spectral edge. We assume for simplicity that we are not in such special 
situations. Otherwise, the divergence of the density of states at the band 
edges modifies the power of $f_L-f$, but does not change the conclusion.   

\subsection{Spontaneous symmetry breaking in cavity arrays}
\label{ch_with_E}

By solving the system of Eqs.\eqref{ch_anom} for the anomalous averages one 
obtains in the steady state
\begin{subequations}
\begin{align}
\alpha_k = & \frac{f-1/2}{f_L-f-i \frac{\Gamma}{4|g|^2}\Delta_q}\, E\,N 
\delta_{k,0} \, ,\\ 
\varphi_k = & 
\frac{\kappa-2i \Delta_q}{2|g|^2} \,\frac{f-1/2}{f_L-f-i 
\frac{\Gamma}{4|g|^2}\Delta_q}\, E\,N \delta_{k,0} \, ,
\end{align}
\end{subequations}
where we used the fact that the photon spectrum is even $\Delta_{-q}=\Delta_q$.
It is clear that the survival of these quantities in the $E\to 0$ limit 
is possible only if the symmetry-breaking field is chosen to excite a 
resonant mode, i.e. $\Delta_q=0$. In this case we get the simpler forms
\begin{subequations}
\begin{align}
\alpha_k = & \frac{f-1/2}{f_L-f}\, E\,N 
\delta_{k,0} \, ,\\ 
\varphi_k = & 
\frac{\kappa}{2|g|^2} \, \frac{f-1/2}{f_L-f}\, E\,N \delta_{k,0} \, ,
\end{align}
\label{ch_symbr_anom}
\end{subequations}
Moreover $f$ has to converge to $f_L$ as the first power of the field. In 
order to prove this statement we now solve Eqs.\eqref{ch_norm} including the 
contribution of the anomalous averages.

By proceeding exactly as in Section \ref{ch_no_E} one obtains the 
symmetry-broken version of Eq.\eqref{lorentz}
\begin{align}
f_N-f = |g|^2 & \frac{1}{2 \pi}\int_{-\pi}^{\pi} \frac{A}{(2\cos 
(k-q)-\Delta)^2+ \delta^2}\, d k  \nonumber \\ 
     +\, & \frac{\kappa}{|g|^2}\, \frac{(f-1/2)^2}{(f_L-f)^2}\, E^2 \, .
\label{E_lorentz}
\end{align}
Again, one can consider $E$ real, without loss of generality. The shift by $q$, 
introduced by the space modulation of the coherent exciting 
field, is irrelevant for the result, since it appears in a periodic function 
integrated over a whole period. Integrating as above one is left with
\begin{align}
f_N-f = |g|^2 & \frac{A}{B\sqrt{f_L-f}} \Re \left 
\{\frac{i}{\zeta_2-\zeta_1}\right \} \nonumber \\
           +\, & \frac{\kappa}{|g|^2}\, \frac{(f-1/2)^2}{(f_L-f)^2}\, E^2 \, .
\label{limits}
\end{align}
It is obvious now that performing first the $E \to 0$ limit is killing both
$\alpha_k$ and $\varphi_k$, since $f$ differs from $f_L$ as long as $g \neq 0$.
After that one is left with the situation described in section \ref{ch_no_E}.

On the contrary, if the scaling limit is taken first, it cancels the first term 
in Eq.\eqref{limits} and with $f \neq f_N$ one concludes that in the 
lasing regime $f_L-f \sim E$, when $E$ goes subsequently to zero. This is 
precisely the condition for the persistence of anomalous averages as the 
symmetry-breaking seed goes to zero. 

\section{Conclusions}

We have proven the instability of the laser systems to symmetry-breaking 
infinitesimal perturbations. In this respect the laser transition is similar to 
equilibrium phase transitions. Due to the invariance with respect to phase 
rotation, the averages like  $\braket{b}$ or $\braket{\sigma}$ have no 
preferred direction in the complex plane and therefore must be zero. 
Non-zero values are generated by a symmetry-breaking seed and we have shown 
that they persist when the seed is removed. Such spontaneous symmetry breaking
takes place when the system is in the lasing phase, and only then.  

We discussed first the case of a single, two-level emitter embedded in a 
cavity and interacting resonantly with its photon mode. Second, we considered a 
chain of such cavities, allowing for photon hopping between nearest neighbors. 
In both cases the symmetry-breaking perturbation was a coherent field $E$, in 
resonant dipole coupling with the exciton.

At the level of the rate equation we showed analytically that the mechanism of 
the spontaneous symmetry breaking is connected to the existence of a sharp
laser transition, with a well-defined threshold point. This involves the 
scaling limit $\kappa \to 0$ and $g \to 0$, so that $g^2/\kappa$ remains finite. 
It was shown that the correct limit order is scaling limit first, followed by $E 
\to 0$. In the case of cavity arrays, the symmetry breaking occurs in the 
resonant Bloch mode, which is also the lasing mode.

Numerical estimates have shown that the sensitivity to symmetry-breaking 
perturbations develops and manifests itself even before reaching these limits.
Anomalous averages remain high if the conditions for lasing are met, but drop 
fast with the perturbation otherwise. In this sense such averages can be used 
as evidence for lasing. For instance, numerical examples showed that a large 
value of $\braket{b}$ confirms the $g^{2}(0) \approx 1$ test. The former has 
the advantage that $\braket{b}$, as a singlet, is accesible at the 
rate-equation level, while $g^{2}(0)$, which contains a quadruplet, is not.    

\section*{Acknowledgment}
The author acknowledges financial support from CNCS-UEFISCDI
Grant No. PN-III-P4-ID-PCE-2016-0221.

\bibliography{symm_br}
\bibstyle{prb}

\end{document}